\documentclass[12pt,preprint2]{aastex}
\usepackage{natbib}\usepackage{txfonts}\usepackage{balance}
\usepackage{graphicx}
\begin{document}
\def\teff{$T\rm_{eff }$}
\def\kms{$\mathrm {km s}^{-1}$}

\title{
Quasometry, Its Use and Purpose
}

\author{
Valeri Makarov, Ciprian Berghea, David Boboltz, Christopher Dieck, Bryan Dorland, Rachel Dudik, Alan Fey, Ralph Gaume, Norbert Zacharias}

\affil{
United States Naval Observatory, Washington, DC 20392, USA}
\email{vvm@usno.navy.mil}

\author{
Xuming Lei, Henrique Schmitt}

\affil{
Computational Physics, Inc., Springfield, VA 22151, USA}

\shortauthors{Makarov et al.}

\shorttitle{Quasometry}

\begin{abstract}
Quasometry is precision measurement of celestial positions and apparent motion
of very distant extragalactic objects, such as quasars, galactic nuclei, and QSOs.
We use this term to identify a specific area of research, the methodology of which
differs from that of general astrometry.
The main purpose of quasometry is to link the sub-milliarcsecond radio
frame (ICRF) with the existing and emerging optical reference frames of similar accuracy,
constructed by astrometric satellites. Some of the main difficulties in achieving this goal
are discussed, e.g., the extended structures of quasar hosts, apparent motion on the sky,
optical variability, galactic companions, faintness. Besides the strategic purpose, quasometry is undoubtedly
useful for global astrometric surveys, as it helps to verify or even correct the
resulting reference frames. There are two options of using measurements of distant quasars
in a global astrometric solution: 1) hard constraints embedded in the fabric of observational equations;
2) {\it a posteriori} fitting of zonal errors. The relative benefits and shortcoming of the two
options are reviewed. A relatively small set of about 200 carefully selected reference quasars can 
go a long way in improving the astrometric value of a space mission, if they are sufficiently 
bright, stable, fairly uniformly distributed on the sky, and are defining sources in the ICRF. 
We present an ongoing program at the USNO to construct a quality set of optical quasars with the required properties and to enhance the ICRF with new sources in the areas where known, well-observed quasars are scarce.
\end{abstract}

\section{Introduction}
The currently adopted International Celestial Reference System (ICRS) is realized through the
second release of the International Celestial Reference Frame\footnote{http://hpiers.obspm.fr/icrs-pc/}
\citep[ICRF2,][]{bob}.
The ICRF2 achieves a stability of the axes of $\simeq 10$ $\mu$as, and an individual object
accuracy down to $\simeq 40$ $\mu$as. To take advantage of this unprecedented accuracy,
a future mission of optical astrometry should directly measure the positions of
ICRF2 sources along with reference and field stars. Depending on the expected intrinsic
accuracy of the mission, these measurements can be used to improve the absolute accuracy
of the resulting optical reference system, or to verify the external accuracy with the
independent data. In this paper, the first possibility is discussed. The previous effort
in constructing an optical analog of the ICRF2 preceding the anticipated global astrometric
missions was directed toward compiling large catalogs of QSOs and AGNs \citep{sou}, counting
more than a hundred thousand objects, and the derivative optical realization of the quasar-based
reference frame \citep[LQRF,][]{and}, obtained by aligning the axes with the ICRS at the milliarcsecond
level. Quasars are by no means ideal objects for optical astrometry. 
Various complications have to be dealt with, e.g., 1) optical variability, 2) extended structures, 
3) apparent motion due to binarity, gravitational lensing, or jets. Even a small admixture of
strongly perturbed sources can nullify all the perceived advantages of a quasar-based reference frame,
while the currently available information about the vast majority of LQRF objects is scarce.
We describe an alternative approach adopted at the USNO of selecting a much smaller set of a few hundred
ICRF2 quasars, carefully vetted and investigated at optical wavelengths. 

\section{Global solutions with constraints}
The estimated performance of a global astrometric mission is usually characterized by a mission-average 
error (e.g., of parallax), which is a very compact, but not so comprehensive, characteristic of
the propagating error. A different approach is to consider the distribution of position-correlated
error (often called {\it zonal} error) on the sky, whether of accidental or systematic origin.
Any discrete distribution of astrometric error on the sphere can be uniquely decomposed into a series of
either vector spherical harmonics (for positions and proper motions) or scalar spherical harmonics
(for parallax). Furthermore, a global astrometric solution can be obtained in terms of the spherical
harmonics rather than for individual objects. Due to the {\it near-orthogonality} of the spherical
harmonics, such a solution is precise for any subset of the basis functions. Using this idea, the accuracy
of a very large global solution can be quickly evaluated on regular computers, such as a standard laptop.
This also allows us to perform a large number of mission simulations for different configurations of
reference objects, single measurement precision, etc. Alternatively, since the eigenvectors of a discrete
zonal error are the spherical harmonics, the covariances and the confidence of a given mission-average
accuracy can be easily computed. The confidence levels are related to the uncertainty of the outcome, 
i.e., Òthe sigma of the sigmaÓ. Following this recipe, the {\it probability distribution} of the mission-average error can be computed without multiple Monte Carlo simulations of the entire mission.
\begin{figure}[t!]
\resizebox{\hsize}{!}{\includegraphics[clip=true,angle=270]{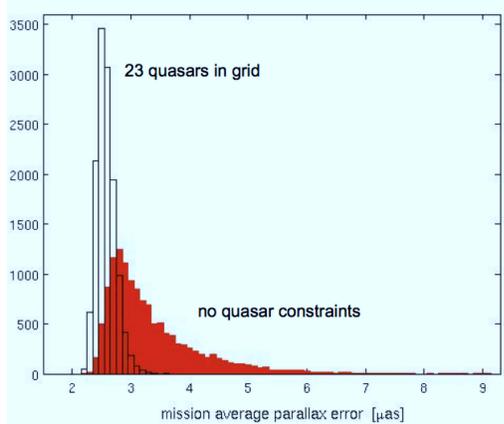}}
\caption{\footnotesize
Distributions of mission-average parallax error of the SIM interferometer project with and without
23 ICRF quasars used as hard constraints in the global solution, but with otherwise the same
input parameters. The remarkable improvement in the expected parallax accuracy and the
tightening of the uncertainty of the mission outcome arises from the reduced propagation of
accidental zonal errors in low-order spherical harmonics.}
\label{distr.fig}
\end{figure}

Figure \ref{distr.fig} shows the distributions of the mission-average parallax error with 23 reference
quasars and without any quasars computed for the Space Interferometry Mission (SIM).
Using only 23 quasars in the grid solution as absolute zero-parallax constraints
not only improves the mission expectancy, but practically eliminates the risk of a very bad outcome. Interestingly, doubling the number of reference
quasars results in a relatively modest improvement of the expectancy of $\simeq 10$\%, and doubling again
in an even smaller improvement. The reason for this weak dependence on the number of quasars is
the ``red" spectrum of accidental zonal error, i.e., that most of the error is carried by the low-order
spherical harmonics.

\begin{figure}[t!]
\resizebox{\hsize}{!}{\includegraphics[clip=true]{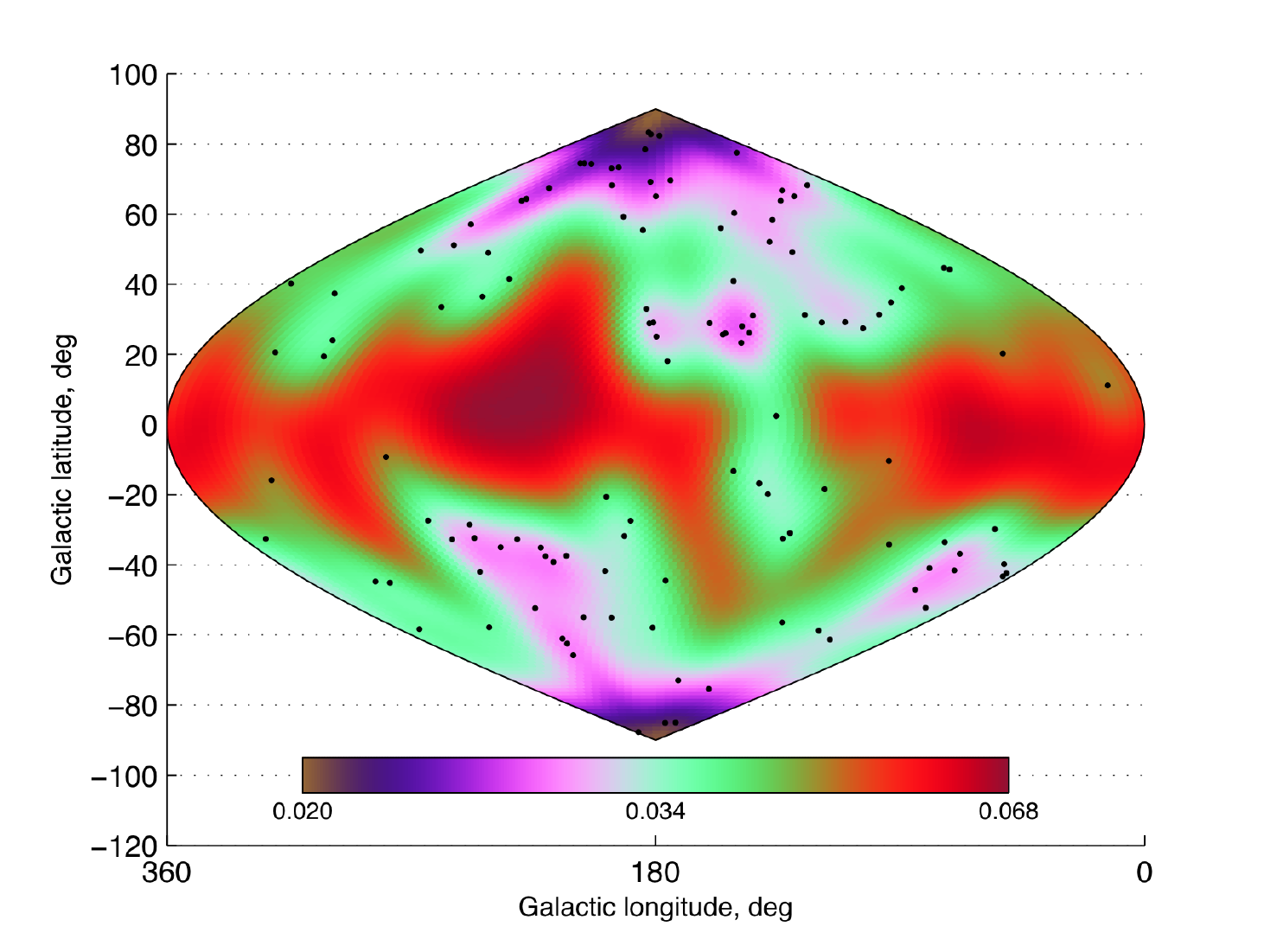}}
\caption{\footnotesize
Distribution of accidental parallax error (standard deviation) simulated for a global astrometric solution
with about 80 grid quasars used as hard constraints in the condition equations. The quasars are shown
with black dots. The error of unit weight is color-coded as shown in the color bar legend below the sky
projection.}
\label{sky.fig}
\end{figure}
To fully realize this advantage, the quasar constraints should be hard-coded in the observational equations.
For example, the ``hard" parallax constraints are implemented by removing the corresponding
columns in the global design matrix. The benefits of such constraints are clearly seen in Fig. \ref{sky.fig}
where the standard deviation of the combined zonal error of parallax for a global astrometric mission is depicted. The error expectancy is much smaller in the areas populated by reference quasars than in the
near-Galactic plane zone where ICRF quasars are not available. An alternative approach is to use
the prior information (such as that the parallax of a quasar is practically zero) in {\it a posteriori} adjustment
of the solution. At the minimum, a constant zero-point of all parallaxes (which is the first spherical
harmonic) can be determined and subtracted. This approach is more cumbersome and less accurate. However,
it may be safer if some of the reference quasars are astrometrically perturbed.

\section{The dangers of quasometry}

\begin{figure}[t!]
\resizebox{\hsize}{!}{\includegraphics[clip=true]{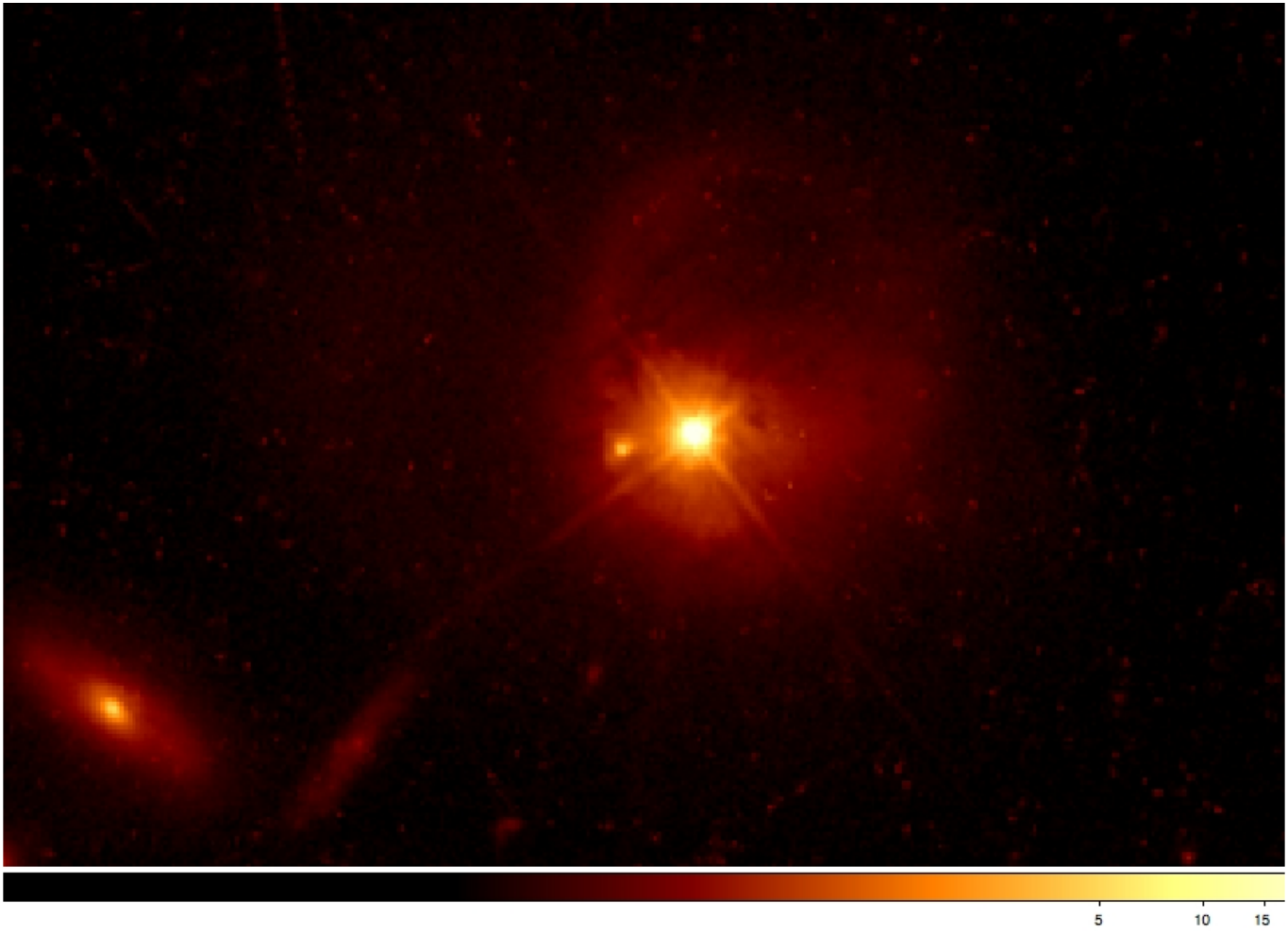}}
\caption{\footnotesize
Archival HST image of the radio-loud quasar PKS 2349-014 at F606W. The map reveals a galaxy companion
east of the optical core, separated by approximately 2 arcsec, and an extended asymmetric structure
with tidal arms.}
\label{PKSo.fig}
\end{figure}
\begin{figure}[t!]
\resizebox{\hsize}{!}{\includegraphics[clip=true]{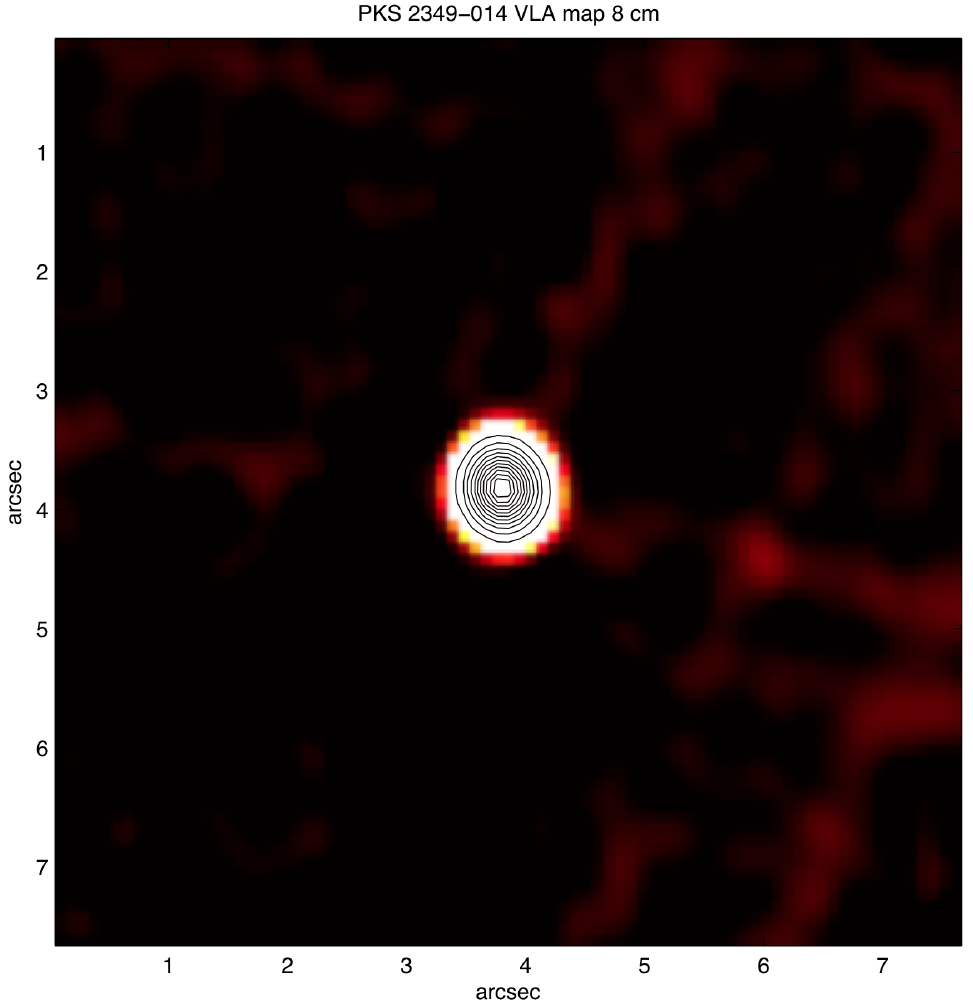}}
\caption{\footnotesize
VLA map of PKS 2349-014 reveals a perfectly unresolved, single radio core.}
\label{PKSr.fig}
\end{figure}

\citet{bah} investigated one of the nearest and brightest quasars, PKS 2349-014. 
Their images obtained with the Hubble Space telescope (HST) with the F606W filter
(yellow-orange light) reveal a complex structure consisting of several components
(Fig. \ref{PKSo.fig}) extending to at least 10 arcseconds. The unresolved nucleus
is the brightest component of this structure, while other components have
estimated magnitudes between 17.9 and 21 (Table \ref{pks.tab}). Even though the
extended components are a few magnitudes fainter than the nucleus, they may
strongly perturb the astrometric determination because of their large extent and
offsets from the nucleus. Additionally, the substrate galaxy is obviously off-center
with respect to the nucleus, and the faint companion galaxy is at a projected separation of
$\simeq 2$ arcsec. These effects will cause a nontrivial photocenter shift. In particular, if we apply
the center-of-mass (CoM) centroiding method to an object of this complexity, disastrous
astrometric errors would ensue, given in the last column of Table \ref{pks.tab}. At the
same time, the VLA map at 8 cm (Fig. \ref{PKSr.fig}) confirms PKS 2349-014 as a
high-quality, unresolved radio core without any extended features or companions. In this case,
a bona fide reference quasar in the radio is a horrible source in the visible light.

\begin{table}
\caption{Magnitudes and predicted astrometric offsets of optical components of PKS 2349-014.
The data in the second column are from \citep{bah}, the data in the third column are derived by us.}
\label{pks.tab}
\begin{center}
\begin{tabular}{lrr}
\hline
\\
Component & $m_{606W}$ & offset  \\
& mag & mas\\
\hline
\\
unresolved nucleus\dotfill  &$ 15.3 $ & $0 $ \\
wisps (tidal tails?)\dotfill   &$ 17.9 $ & $150 $\\
extended nebulosity \dotfill   &$ 18.0 $ & $200 $\\
companion galaxy \dotfill  &$ 21.0 $ & $10 $\\
host galaxy centered on nucleus \dotfill &$ 18.6 $ & $0 $ \\
\hline
\end{tabular}
\end{center}
\end{table}
Using more distant quasars as reference objects for an optical frame will certainly
alleviate the problem, but only to some extent. A quasar like the PKS 2349-014, but
100 times more distant (which would be farther out than the most distant known quasar), 
would still be marginally perturbed for a mission like Gaia.
The situation is drastically better if a more sophisticated method of image centering
than CoM is used. \citet{guy} performed a NIR adaptive optics imaging survey of nearby
($z<0.3$) QSOs. Using the superb resolution of their images, they fitted the template
point spread function (PSF) to all observed profiles. Estimating the prospects of quasometry,
the important result is that most of the images are quite close to the PSF in the
core areas of approximately 1 arcsec radius. As the majority of luminous QSOs are associated
with either ellipticals or mergers, the host galaxies and other extended features become
prominent, and eventually dominating, outside the cores. Therefore, using specially adjusted
PSF centroiding methods restricted to the core signal should greatly mitigate the impact of
extended features for many (but not all) reference quasars. 

\citet{cam} find a scatter of optical minus ICRF positions up to 80 mas, which seems correlated 
with the X-band structure index. Their result implies that the radio loud quasars can not be used
as reference objects in the optical wavelengths because of the perturbations in their photocenters
from bright extended features. The X-band VLBI images of most of the sources in their study
show that the entire extent of the radio emission is typically less than 5-10 mas. Assuming the optical and radio emission are associated with the same active nucleus, from the astrophysical prospective, it is difficult to understand the nature of the large optical-VLBI offsets reported in this paper. In our opinion, the most likely explanation for the large
scatter is the local systematic error of the UCAC3 catalog \citep{za}, 
which was not taken into account in that
paper. On a scale of 1 to 2 deg on the sky, systematic errors of proper motions for stars in the 13.5 to 16
mag range are caused by the unmodelled systematic errors of the early photographic plates,
which have scales of about $5\degr$ on a side. In particular, for the area  north of $-20\degr$,
UCAC3 relies on Schmidt plate data, and systematic errors of proper motions up to 12 mas yr$^{-1}$ have
been found \citep{roe}. Combined
with the epoch difference of about 8 years between the ESO, SOAR data from 2006-2008 and the UCAC
mean epoch of 1998-2003, this gives rise to a systematic error of position of up to 100 mas,
which will not decrease with the number of reference stars. For the area of the sky well covered
by the SPM early epoch data \citep{gir}, the systematic errors in UCAC3 proper motions are estimated
to be much smaller, but the epoch difference is larger (about 10 years), which could lead to
20 to 40 mas positional offsets not taken into account in the paper by \citet{cam}.
Similarly large offsets between the optical and radio positions of some ICRF quasars were found by \citep{z},
where the possibility of large zonal errors in the optical reference system was discussed.

An additional considerable complication arises from the optical variability of quasars. \citet{tar}
determined the light curves of 41 randomly selected QSOs in a single deep field. During more than 50 months of observations, most of the objects varied by up to 1 mag. The light curves in the $G$, $R$ and $I$ bands
were strongly correlated, and for one of the investigated quasars, a correlation between the
photometric variability and the astrometric location was detected. Such correlated astrometric-photometric
perturbations are to be expected if the photocenter of the image and the estimated centroid is
affected by the external structures, such as the host galaxy, because the extended components of
the object are hardly variable. The situation is much better at radio wavelengths, where the extended
features and inactive companion galaxies are not visible. \citet{fom} investigated four ICRF quasars
and found that some ICRF positions are dominated by a moving jet component and may be displaced up to 0.5 mas, reflecting the motion of the jet. This probably sets the scale of the {\it intrinsic} apparent motion
of the active nuclei. \citet{tit} determined ``proper motions", i.e., linear trajectories of 555 VLBI
radio sources over some 20 years of observation. Even though some of the quasars seem to have motions
of up to 1 mas yr$^{-1}$, the systematic (correlated) part of the vector field is much smaller.
Using their data, we redetermined the vector spherical harmonics up to order 3 for this set
of sources, and set an upper bound on the amplitude of the systematic vector field of 50 $\mu$as $^{-1}$.
None of the fitted spherical harmonics were statistically significant at the $3\sigma$ level.

\section{A precious set}
\begin{figure}[t!]
\resizebox{\hsize}{!}{\includegraphics[clip=true]{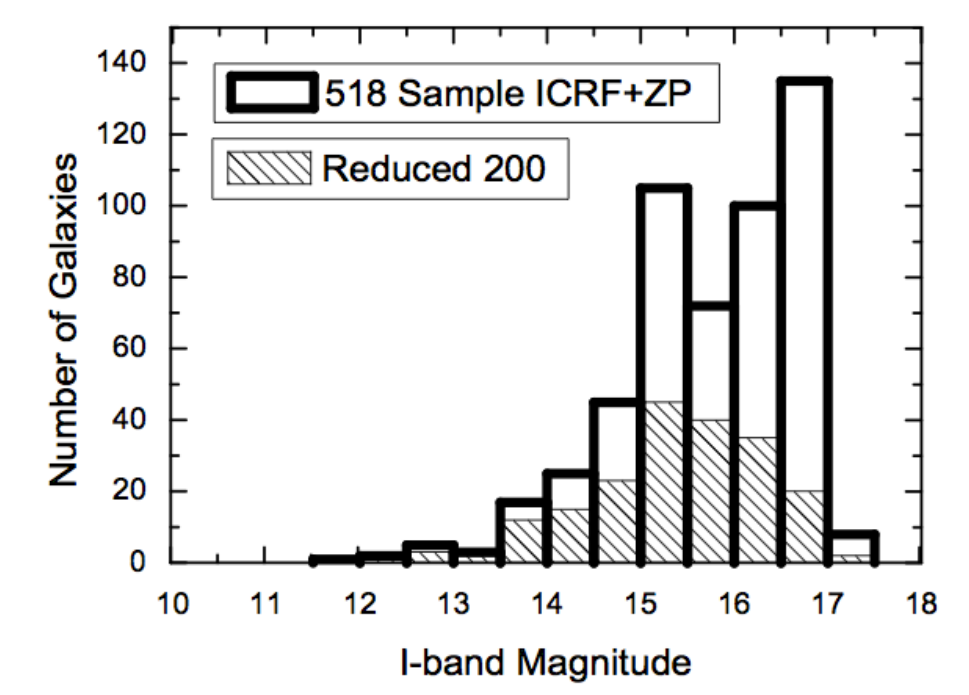}}
\caption{\footnotesize
Distribution of $I$-magnitudes of selected high-quality reference QSOs.}
\label{dudik.fig}
\end{figure}

\begin{figure*}[t]
\resizebox{\hsize}{!}{\includegraphics[clip=true,angle=270]{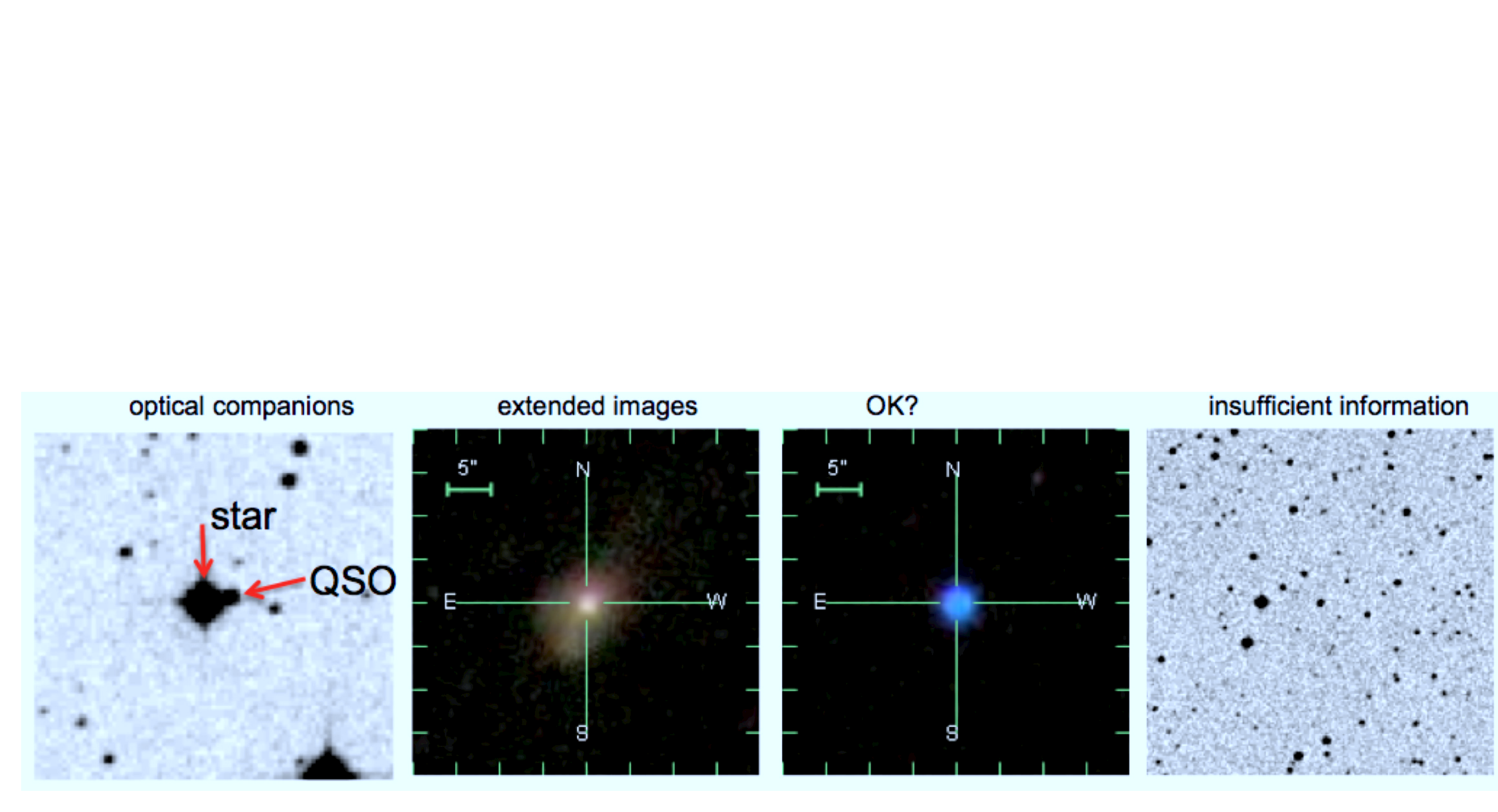}}
\caption{
\footnotesize
Types of ICRF quasars encountered in the process of selecting reference objects
for precision optical astrometry.}
\label{types.fig}
\end{figure*}
The best objects for quasometry are quasars that are both radio loud and optically bright.
Furthermore, the source of radio emission, which mostly comes from the relativistic jet, should be compact
and preferably without superluminal motion. The source of optical emission in the core, which,
to a large extent, is generated by the accretion disk around the central black hole, should be
compact and stable. The host galaxies should be faint and symmetric around the nucleus. 
These criteria seem to contradict each other. According to \citet{sik},
there exists a dichotomy of accretion luminosity versus radio loudness. They surmise that
the black holes in ellipticals rotate faster, which leads to optically brighter nuclei.
However, AGNs become radio quiet at very high rates of rotation. Presumably, high rates of rotation preclude collimation of jets. This perhaps accounts for the sad tendency of most accurate ICRF sources to be
optically faint. 

\citet{ham} infer a ``fundamental plane" of QSOs on a sample of 70 objects with $0.06\le z \le 0.46$
observed with HST's WFPC2. This relation can be re-written as
\begin{eqnarray}
M_V({\rm host})-M_V({\rm nuc})&=&-0.697 M_V({\rm nuc})-15.9 \nonumber\\
&&
\end{eqnarray}
which includes the absolute $V$ magnitudes of the nucleus and the host galaxy. According to
this relation, the magnitude difference is statistically larger for more luminous nuclei, i.e.,
the intrinsically bright cores should be much less perturbed by the host galaxies. The host and the nucleus
have equal magnitudes at $M_V({\rm nuc})=-22.8$, but the difference becomes 2 mag at $M_V({\rm nuc})=-25.7$.
In this respect, the exceptionally luminous nearby quasar 3C 273 at $\Delta M_V=3$ is a good object for quasometry, but PKS 2349-014 at $\Delta M_V=-0.2$ is bad.

Many of the strategic purposes of optical quasometry can be fulfilled with a relatively small set
of quasars of prime quality. For example, the alignment of the rotation and spin of an
optical reference frame with the ICRS does not require a large number of common sources, but each
source should be consistent at the optical and radio wavelengths to better than $\sim 1$ mas.
We believe that selecting a limited sample of quality objects is a more promising approach than
using as many unvetted sources as possible. An ongoing activity at the USNO is addressed at
vetting at least 200 QSOs that can serve as the foundation of the future radio-optical reference frame.
A parallel activity is under way in Europe \citep{bour}.

We are using the following selection criteria:
\begin{itemize}
\item Preferably high-accuracy ICFR2 source
\item Brighter than $I\simeq 17$ mag
\item Photometrically stable, $\delta I<0.5$
\item Not a merger 
\item Not too close, $z \geq 0.1$
\item No nearby projections of stars or galaxies
\item Nucleus is significantly brighter than the host galaxy.
\item Uniformly distributed on the sky
\end{itemize}
The $I$-band magnitude distribution of the original selection of 518 ICRF and radio-mute QSOs
is shown in Fig. \ref{dudik.fig}. A second cut at this sample requires additional investigation
of individual objects involving archival data and new observations. In particular, the photometric
campaign commenced in 2005 from both hemispheres \citep{ojh} goes on. It already resulted in the
identification of highly variable objects, and of quasars with erroneous magnitudes in the catalogs.

An extensive search for and perusal of images in the literature and in the digitized archives
resulted in elimination of further several dozen objects. Some of the typical cases we have encountered in
this process are depicted in Fig. \ref{types.fig}. Some of the radio-loud quasars are projected too close to
a bright star, others feature bright extended components or merger structures. Still, for many of the
candidate reference QSOs, the available information is too scant for making a final decision. New
good-seeing or AO images on large telescopes are needed for about a hundred of the original objects.

\begin{figure}[t]
\resizebox{\hsize}{!}{\includegraphics[clip=true]{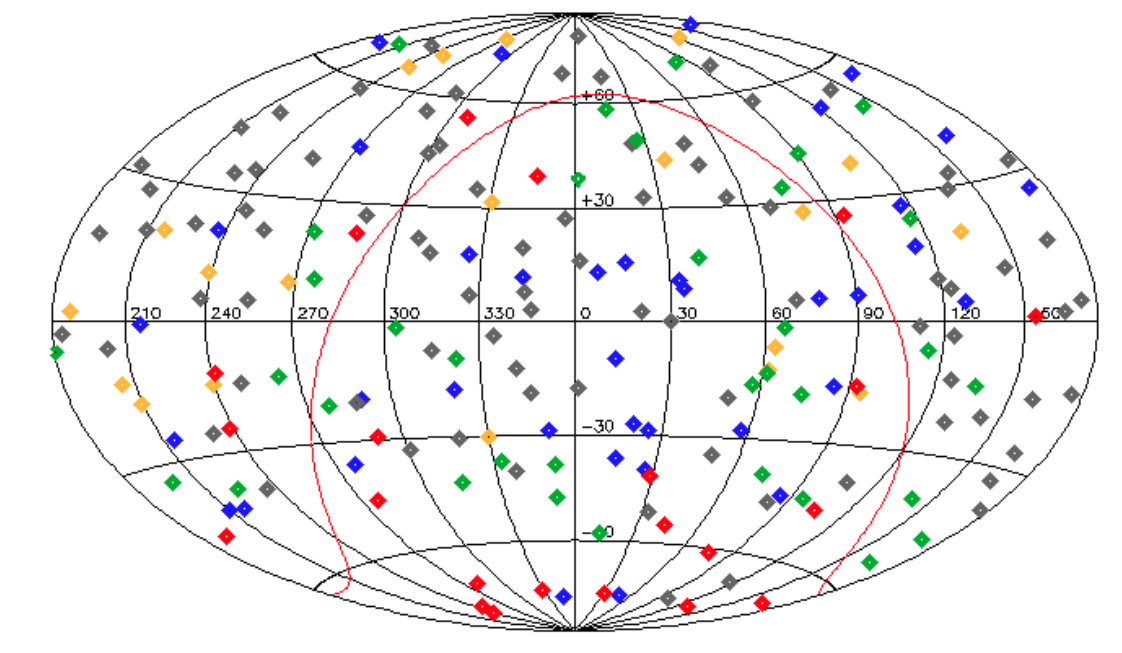}}
\caption{
\footnotesize
Distribution of defining ICRF quasars vetted for the future optical reference frame. Magnitudes:
yellow = 12-14, blue = 14-15, green = 15-16, black = 16-16.5, red = 16.5-17 mag.}
\label{mags.fig}
\end{figure}
Last but not least, the reference grid quasars should be as uniformly distributed on the sky as possible.
The degree of ``emptiness" of the area around a given quasar is a significant figure of merit
in our selection. To quantify this figure of merit, we developed a {\it directional near neighbor
statistic}, which is more suitable for this task than the traditional near-neighbor statistic
(the distance to the nearest neighbor). If an object sits on the edge of a large void, such as seen
in the lower left part of the sky projection in Fig. \ref{mags.fig} around the Galactic plane,
it should be given a high figure of merit even if another objects happens to lie close to it
outside the void. The directional near neighbor (DNN) statistic is the mean of 12 angular distances to
the nearest neighbors within 12 directional segments of equal width. More complicated rank orders
can be constructed by combining the DNN statistics with magnitudes, since a strong preference is
given to brighter quasars in our selection.

\acknowledgements
We thank the USNO Editorial Board for helpful suggestions and
a critical reading of the original version of the
paper.

\end{document}